\documentclass{aa}  

\usepackage{graphicx}
\usepackage{xspace}
\usepackage{txfonts}
\usepackage{subfig}
\usepackage{xcolor}

\begin{document}

   \title{Understanding the unusual life of the Cartwheel galaxy using stellar populations}

 \author{F. R. Ditrani
          \inst{1,}\inst{2}, M. Longhetti\inst{2}, M. Fossati\inst{1,2}, A. Wolter\inst{2} }
    \institute{Università degli studi di Milano-Bicocca, Piazza della scienza 3, I-20126 Milano, Italy\\
   \email{f.ditrani1@campus.unimib.it}
   \and
   INAF-Osservatorio Astronomico di Brera, via Brera 28, I-20121 Milano, Italy\\}

   \date{Received; accepted}
 \titlerunning {Understanding the unusual life of the Cartwheel galaxy}
 \authorrunning{Ditrani et al.}
 
\abstract
   {Collisional ring galaxies (RiGs) are the result of the impact between two galaxies, with one of them passing close to the centre of the other, piercing its gaseous and stellar disc. In this framework, the impact generates a shock wave front that propagates within the disc of the target galaxy soon after the encounter, producing a characteristic expanding ring-shaped structure. RiGs represent one of the most extreme environments in which we can study the physical properties of galaxies and the transformations they undergo during collisions.
   
   The paradigm RiG is the Cartwheel galaxy at $z = 0.03$. This galaxy has been the object of both theoretical and observational studies, but the details of the mechanisms that lead to its peculiar morphology of double rings with connecting spokes and to its physical properties are still far from clear.
   
   To shed light on the history of the Cartwheel galaxy, we performed a spatially resolved analysis as a function of galactocentric distance, exploiting spectroscopic data from VLT/MUSE observations combined with photometric data covering a large wavelength range, from the UV GALEX to the IR JWST/MIRI. Using full-index fitting of the stellar spectra, an analysis of the nebular emission, and joint full spectral and photometry fitting, we derived stellar ages, gas and stellar metallicities, and star formation histories (SFHs) in four spatially distinct regions of the galaxy.
   
   We find that, apart from the peculiar morphology, a large fraction of the Cartwheel galaxy is not affected by the recent impact from the companion bullet, and retains the characteristics of a typical spiral galaxy. On the contrary, the outer ring is strongly affected by the recent impact, and is completely dominated by stars formed not earlier than $\sim 400$ Myr ago. Our picture suggests that the collision shock wave, while moving forward to the external region of the galaxy, drags the already formed stars, sweeping the inner areas outwards, as proposed by recent collision models. At the same time, the ages found in the external ring are older than the predicted timescale of the ring expansion after the collision.}
   
   
   


   \maketitle
%

\section{Introduction}

Collisional ring galaxies \citep[hereafter RiGs; see e.g.][]{Athanassoula1985} are a peculiar class of galaxy thought to have suffered violent interaction events in the recent past. They appear to be the result of the impact between two galaxies, with the smaller one passing through the disc of the larger one ---which is referred to as the target galaxy--- close to its centre. In this framework, a shock wave front induced by the impact starts to propagate within the disc of the target galaxy soon after the encounter, producing the characteristic expanding external ring \citep[e.g.][]{Lynds1976,Theys1977,Hernquist1993,Mihos1994,HorCombes2001,Mapelli2008a,Map2008}. 
The high values of star formation rate (SFR) measured within the rings \citep[$18-20$ M$_\odot$ yr$^{-1}$, see][]{Appleton97,Wolter2004,mayya2005} have always been considered to be associated with the shock waves and to be responsible for triggering the events.

RiGs represent one of the most extreme environments in which we can study star formation, not only for the energetic contribution from the encounter, but also due to a relatively `clean' population of spatially segregated regions, which can be studied individually. Many collision models have been developed to describe the details of the mechanism of RiG formation \citep[see][and references therein]{Mapelli2008a}, and all of them depict the advancing front as leaving behind gas  and recently formed stars while triggering new generations of stars at the new position of the expanding ring. Only recently, \cite{renaud2018} proposed a new model in which the wave front, while expanding towards the external region of the target galaxy, drags gas and recently formed stars, sweeping the inner areas, that is, moving the stars and gas outwards.

The most famous RiG is the Cartwheel galaxy, a member of the compact group SGC 0035$-$3357 in the southern hemisphere \citep[][]{Iovino2002AJ} at redshift $z = 0.03$ \citep{amram1998}, corresponding to a luminosity distance of $131$ Mpc. 
Among the RiGs, and generally among all galaxies, the Cartwheel galaxy hosts the highest number of bright ultraluminous X-ray sources (ULXs). 
ULXs are extra-galactic, point-like, non-nuclear X-ray sources with 10$^{39}$ erg/s $<$ L$_{X}$ $<$ 10$^{42}$ erg/s \citep[see][for reviews]{Kaaret17,King2023} and are often associated with recent star formation activity \citep{mineo2012}. Indeed, ULXs can be the end products  of (possibly very short-lived) star formation episodes. It has also been proposed that the observed excess of ULXs in RiGs is driven by the low gas-phase metallicity of their galactic environment \citep[e.g.][]{mapelli2009}, but statistically significant estimates of the nebular metallicities are scarce and many of them have 
only recently become available \citep[][]{bransford1998A,Kosta2021,zaragoza2022nebular}.

The Cartwheel galaxy has been the object of many studies, both theoretical and observational, but the details of the mechanisms that lead to its peculiar morphology and its physical properties are still far from clear.
More specifically, the age of the collision, a critical quantity in order to understand the evolution and dynamics of the galaxy, is not easy to establish: \cite{higdon1996}, analysing the HI velocity fields, inferred that $300$ Myr have passed since the impact; \cite{amram1998} estimated the age of the ring at $>200$ Myr from the H${\alpha}$ kinematics ($13-30$ km s$^{-1}$ expansion).
\cite{struck1996} further suggested that the expanding external ring loses material that moves back towards the nucleus along the spokes, triggering star formation events.
The simulations of \cite{renaud2018} follow a similar dynamical scenario, but with a shorter timescale ($<100$ Myr) for the persistence of the ring.
Recently, \cite{zaragoza2022nebular} analysed the oxygen abundance in the HII regions populating the external ring of the Cartwheel galaxy, and found that it is lower by $\approx 0.1$ dex with respect to the extrapolation of the inner radial gradient. This observed evidence has been interpreted as due to the effect of dragging by the shock wave that leaves the inner regions of the galaxy almost undisturbed, while displacing both the stars formed during the collision and the potential metal-poor gas from the bullet galaxy into the structure that we recognise as the external ring.
The recent work of \cite{MayyaAge}, who analysed Astrosat/UVIT far ultraviolet (FUV) imaging data of the Cartwheel galaxy, revealed evidence of a wide range of ages of the stellar populations in the HII regions of the external ring, supporting the collision model of  \cite{renaud2018}.

The collision scenarios can be strongly discriminated by means of spatially resolved studies of the stellar and nebular content in RiGs. Indeed, the stellar and nebular properties in galaxies and their spatial distribution contain the signatures of the physical processes that lead to their formation and present morphology. Mergers and interactions among galaxies have the effect of mixing different stellar populations and producing gas inflow and outflow, which can trigger star formation activity \citep[e.g.][]{Hopkins2013,Ferreras2017}.
Studies of stellar population properties in galaxies have proven very successful in deriving information on the formation and evolution history of galaxies, providing insights into the physical mechanisms of the processes involved \citep[e.g.][]{Thomas2011,Conroy_rew_2013,Maraston2020}.
The age, metallicity, and chemical composition of the stellar content in galaxies can be estimated by measuring the equivalent width of well-defined spectral features \citep[the so-called narrowband spectral indices; see][]{worthey1994old,Thomas2003,Johansson2012} or by exploiting the full spectral fitting technique \citep[e.g.][]{CidFernandes2005,cappellari2017improving}. Recently, a hybrid approach has been proposed consisting in fitting only the specific wavelength regions involved in well-known spectral indices  pixel
by pixel \citep[][]{martin2019fornax}.

In the present paper, we present our analysis of the stellar populations of the Cartwheel galaxy with the aim of deriving a global picture of the recent impact. Differently from the previously published analyses, this work, for the first time, considers multi-wavelength data for the whole galaxy and is not restricted to some specific peculiar components (e.g. HII regions). Our aim is to represent the total stellar content distribution of the Cartwheel galaxy in detail.  Another novelty of this work is in our use of the recently available infrared JWST Early Release Observation photometric data, which have proven to be a fundamental tool for detecting old, low-mass stars and therefore the star formation history (SFH) within this peculiar galaxy. 
In a forthcoming paper, we will discuss the relation between the physical properties of the star forming regions in the Cartwheel galaxy and the characteristics of its ULX population.

In Sect.~\ref{section:datadata} we present the full set of data used to perform our analysis and the definitions of the different regions within the Cartwheel galaxy. In Sect.~\ref{section:specextract} we describe in particular how we derive the co-added optical spectra for each region and the procedures adopted to disentangle the stellar from the nebular components. Sect.~\ref{section:analysis} details the analysis performed both on the stellar and nebular components, and our results are presented in Sect.~\ref{section:results}.  In Sect.~\ref{section:finaldiscussion} we present a discussion of our analysis.
Throughout the paper, we adopt a standard $\Lambda$CDM cosmology with $\Omega_M = 0.286$, $\Omega_\Lambda = 0.714,$ and $H_0 = 69.6$ km s$^{-1}$ Mpc$^{-1}$ \citep{wright2006cosmology,bennett20141}. Magnitudes are in the AB system \citep{oke1974absolute}.

\section{Observational data and region identification}
\label{section:datadata}
\begin{figure*}
   \centering
   \includegraphics[width=0.95\textwidth]{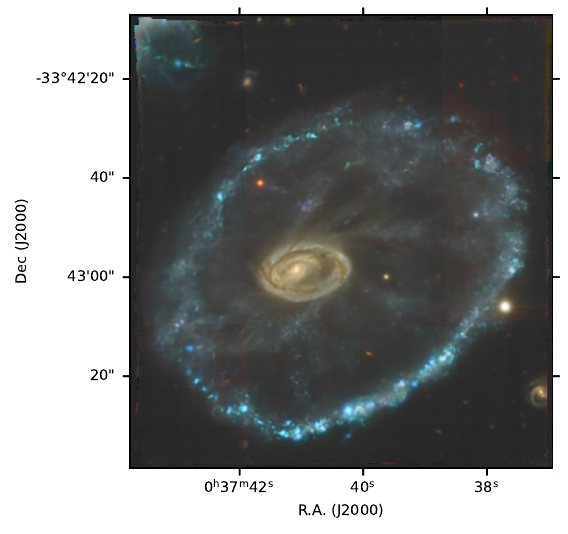}
      \caption{False-colour image of the Cartwheel galaxy from MUSE data. Each RGB channel was constructed from the reduced MUSE data cube by projecting it in the wavelength ranges $4800-6100\AA$, $6100-7500\AA$, and $7500-9000\AA$ for the B, G, and R channels respectively. The image covers a field of view of $\approx 2 \times 2$ arcmin equivalent to nearly $74$ kpc on a side at the Cartwheel's redshift.}
         \label{fig:cartwheel}
   \end{figure*}

We used integral field spectroscopic data of the Cartwheel galaxy from the Multi Unit Spectroscopic Explore (MUSE) mounted on UT4 of the ESO Very Large Telescope. Four mosaicked observations have been carried out during the MUSE science verification run in August $2014$ (program ID: 60.A-9333) and cover a field of view (FoV) of $2\times2$ arcmin, including the entire ring ($1.4 \times 1.5$ arcmin). These data have a total exposure time of $8400$ s. We downloaded the fully reduced 3D science data cube from the ESO Science Portal. The data cover the wavelength range from $4750$ to $9351$ $\AA$, with a spatial sampling of $0.2$ arcsec/pixel, a spectral dispersion of $1.25 \AA$ and a spectral resolution of $\sim 3$ $\AA$ \citep{bacon2010muse}. Figure~\ref{fig:cartwheel} shows a colour image of the Cartwheel galaxy obtained using the MUSE data cube and considering three wavelength ranges as pseudo photometric bands: $4800-6100\AA$ as {B} channel, $6100-7500\AA$  as {G} channel, and $7500-9000\AA$ as {R} channel.

We also exploited the available photometric data for this galaxy: FUV and NUV imaging from the Galaxy Evolution Explorer \citep[GALEX,][]{martin2005galaxy} program GI1-045002, OmegaCAM-VST $u$-, $g$-, $r$-, $i$- and $z$-band images from the public survey \citep{kuijken2011omegacam}, the JWST-NIRCam F090W, F150W, F200W, F277W, F356W and F444W images, and the JWST-MIRI F770W, F1000W, F1280W and F1800W images from the JWST Early Release Observation $1$ (ERO PID 2727, PI Pontoppidan).

In this work we are interested in deriving the stellar population properties in different regions of the Cartwheel galaxy, with the aim of gaining insights on the formation mechanisms which shaped its complex morphology. To address this aim, we started to identify specific spatial regions on the basis of the distribution of the star formation activity. 
We extracted the flux maps of the H$\alpha$ and [NII]6548-6583$\AA$ emission lines from the data cube using the KUBEVIZ code (v2.0).  We refer the readers to \citet[][]{fossati2016muse} for the details of the emission line fitting procedure. The code fits the H$\alpha$+[NII] emission line complex with three Gaussians which share a single recessional velocity and velocity dispersion as free parameters. The flux ratio of the [NII]6548-6583$\AA$ emission lines is fixed to the values from atomic physics given in \cite{Storey2000}. The code first fits all pixels in the MUSE datacube independently. Individual pixels where the $S/N$ of the H$\alpha$ line is less than 5 are then masked. Masked pixels which are adjacent to unmasked ones are fit again, this time using the average values from the fits of nearby good pixels as initial guesses. The code starts from bad fits that are surrounded by the highest number of good fits and iteratively re-evaluates the masking, gradually moving to more isolated masked fits until no further improvement to the spatial flux map is found \citep[see][for more details on the iterative procedure]{fossati2019muse}.

\begin{figure}
   \centering
   \includegraphics[width=0.5\textwidth]{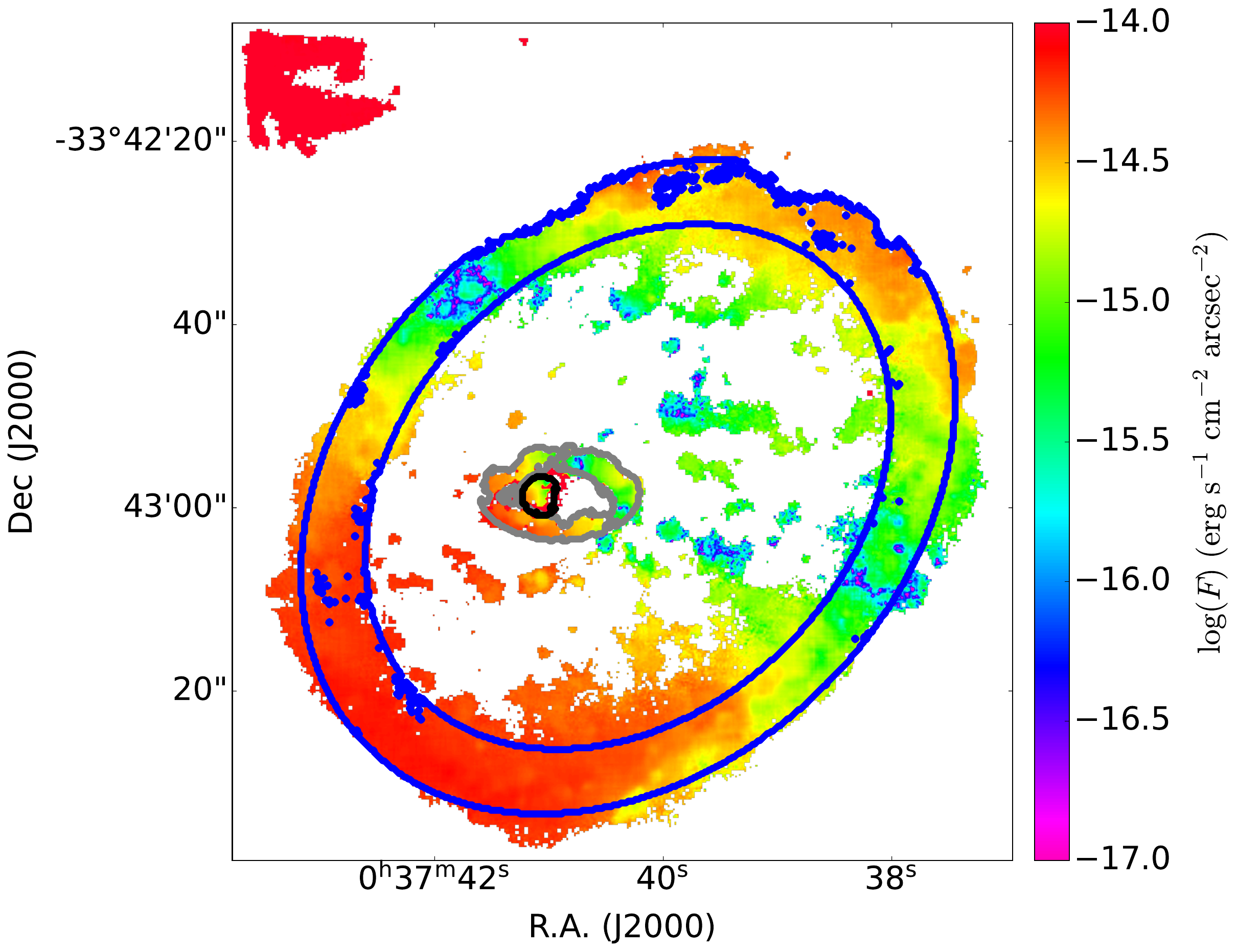}
      \caption{MUSE H$\alpha$ map of the Cartwheel galaxy, where the coloured pixels correspond to S/N(H$\alpha$) $ > 5$ and with kinematic uncertainties on the strongest line lower than $50 \ \text{km s}^{-1}$. The black, grey, and blue contours indicates the {`nucleus'}, {`inner ring'} and {`outer ring'} regions, respectively. The {`in-between'} region is defined by the outer grey contour and the inner blue contour.}
              
         \label{fig:cartwheelha}
   \end{figure}

Figure~\ref{fig:cartwheelha} shows the MUSE map of the H$\alpha$ line flux for pixels with $S/N>5$ for the Cartwheel galaxy.
We readily noticed that the H$\alpha$ distribution is structured in at least three separate spatial regions: a nuclear region (hereafter nucleus, black line), an inner ring (grey line), and an {outer ring} (in blue). We also defined a region between the {inner ring} and the {outer ring} (hereafter the {in-between} region).

\section{Spectral extraction: Stellar continuum and nebular emission}
\label{section:specextract}

From the four spatial regions identified on the H$\alpha$ map of the galaxy, we selected the spaxels with $S/N > 5$ in H$\alpha$ (neglecting the corresponding stellar absorption) and with uncertainties of the H$\alpha$+[NII]6548-6583$\AA$  kinematics lower than $50 \ \text{km s}^{-1}$. 
We extracted the spectra from each of the selected spaxels, and shifted them to rest-frame using the recessional velocity estimate derived from the H$\alpha$+[NII]6548-6583$\AA$ lines. After that, we shifted to the same common rest-frame also the spectra extracted from the spaxels with $S/N < 5$ in H$\alpha$, for which no reliable kinematic information could be obtained, by interpolating a 2D model of the recession velocity field defined by the significant emission lines.

Our investigation is a three-step analysis, where the first two consist of separate analyses of the stellar and nebular components of the four spatial regions of the galaxy, while the third is a joint fit of all the available photometric and spectroscopic data and is partly based on the results obtained in the first two steps.
For the analysis of the first step, we obtained the co-added spectrum of each spatial region excluding from the sum spectra coming from the spaxels within the HII regions identified in  \cite{zaragoza2022nebular}, in order to limit the overwhelming nebular emission in those regions. Indeed, the aim of this part of the analysis is to get insights on the mean properties of the  stellar content in each of the four regions of the Cartwheel galaxy.  The nebular emission lines which are clearly visible in all the four regions are certainly due to highly ionising stars which are mainly confined within the brighter HII regions and which have been dated in a recent paper by \cite{MayyaAge}. Differently from their study, the aim of our analysis of the stellar spectra is to determine the mix of stellar populations characterizing each region in the Cartwheel galaxy, including both the precollisional component and stars possibly formed in the inner part of the galaxy when the shock wave was in more backwards positions. 
Stars within HII regions are so young and bright that would bias the results dominating the total light. The exclusion of the HII regions allows us then to study in a more reliable way the mean stellar population parameters of the Cartwheel galaxy, even in regions that would otherwise be dominated by nebular emission. Consistently with \cite{zaragoza2022nebular}, we modelled each HII region with a circular region of radius $0.6$ arcsec, equivalent to $370$ pc, and we excluded them.

Differently from this, for the further two analysis steps, we summed fluxes from all the available spaxels in order to obtain the co-added spectra representative of the overall emission of each of the four spatial regions in the Cartwheel galaxy. In the outer ring, we also extracted a further co-added spectrum corresponding to the HII regions alone, for which we performed only the nebular component analysis for comparison with other literature studies.

To improve upon the statistical significance of the results and given the large spatial dimensions of the selected regions, we divided the inner ring in four equally extended subregions, the in-between region in eight subregions, and the outer ring in $16$ subregions, while we keep a single spectrum as representative of the nucleus.

We based the analysis of the stellar component on the rest-frame wavelengths between $4650 \ \AA$ and $5400 \ \AA$, the range that is most sensitive to the age and metallicity of the stars. For the analysis of the nebular components and for the full spectral fitting one, we considered spectra over a larger wavelength range, between $4650 \ \AA$ and $7200 \ \AA$ restframe, including both H$\beta$ and H$\alpha$ lines.
For each co-added spectrum, we then separated the stellar and nebular components. To this aim, we first estimated the corresponding stellar kinematics (i.e. residual redshift and velocity dispersion). Since different wavelength ranges receive light contributions from different stellar components, which can be characterised by slightly different kinematics, we performed the kinematic analysis using the absorption features in the range between $4650 \ \AA$ and $5400 \ \AA$, which is the spectral region used for the stellar analysis.
We obtained the residual redshift and velocity dispersion on the co-added spectra exploiting the latest version of the pPXF (Penalized PiXel-Fitting) code \citep{cappellari2004parametric,cappellari2017improving,cappellari2023}. We adopted the MILES stellar spectral library \citep{sanchez2006medium}, convolved to the MUSE instrumental resolution, as model template library.

\begin{figure}
   \centering
   \includegraphics[width=0.5\textwidth]{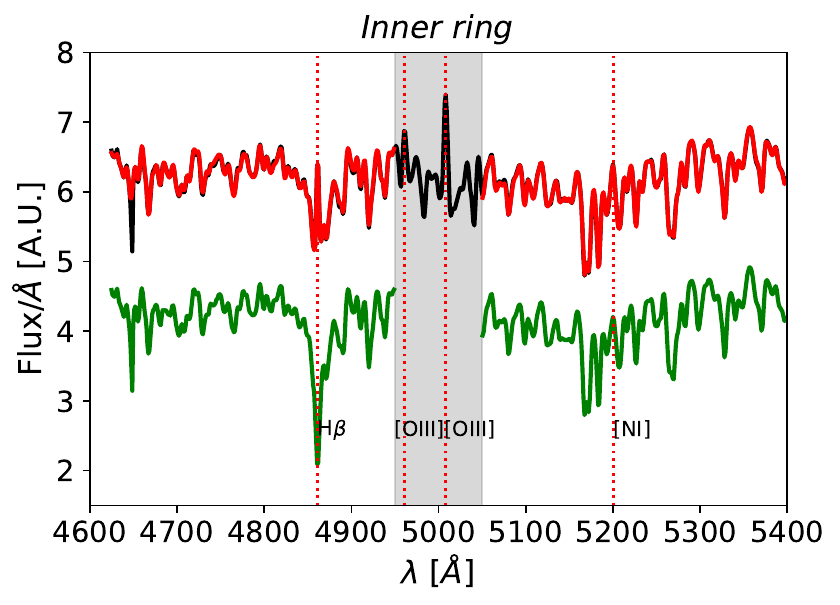}
      \caption{pPXF fit for the spectrum of one subregion of the inner ring. The black line marks the observed spectrum, the superimposed red line is the pPXF best-fit, and the green line is the observed spectrum to which the emission line are subtracted, shifted for clarity. The grey region is masked in the pPXF fit.}
         \label{fig:exsubneb}
   \end{figure}

 We then decoupled the stellar component from the nebular one in each spectrum, running again the pPXF code, this time fixing the kinematics measured in the previous stage. In this case, we used the E-MILES simple stellar population (SSP) synthesis models \citep{vazdekis2016uv}, obtained assuming the BaSTI tracks \citep{pietrinferni2004large,pietrinferni2006large} and \cite{chabrier2003galactic} IMF.  
We fitted the nebular emission lines assuming $6$ moments of their kinematic, since 
we tested that the use of less moments (e.g. a simple Gaussian fit) limits the fit quality, especially for strong emission lines, and produces a residual component affecting the extracted stellar continuum spectra.  Finally, we subtracted the best-fit nebular components from the total spectra, deriving clean stellar component spectra for each of the four regions.
Figure~\ref{fig:exsubneb} shows an example of the extracted stellar spectrum of the inner ring using pPXF. The pPXF code was able, for each spectrum, to perform an accurate fit of the stellar component, disentangled from the emission nebular one.

\section{Analysis}
\label{section:analysis}
In this section, we describe the details of the analysis performed on the four identified regions of the Cartwheel galaxy. As mentioned in Sect.~\ref{section:specextract}, we divided the analysis into three separate steps. The first one is the analysis of the stellar components (Sect.~\ref{subsection:stellarcont} and~\ref{subsection:hierarc}). The second step is the analysis performed on the nebular emission of each of the four regions (Sect.~\ref{subsection:nebular}), and the last step is the joint photometric and spectroscopic analysis aimed at deriving the SFH of the four regions on the basis of the information derived with the previous methods (Sect.~\ref{subsection:mcspf}).

\subsection{Stellar population properties via full-index fitting}
\label{subsection:stellarcont}
We prepared a set of comparison templates based on the SSP models by \cite{vazdekis2016uv}, with the aim of extracting information on the stellar age and metallicity in the four regions of the Cartwheel galaxy from their stellar spectra. Models adopt the BaSTI tracks \citep{pietrinferni2004large,pietrinferni2006large} combined with the MILES \citep[][$2.51\AA$ FWHM resolution]{sanchez2006medium,falcon2011updated} empirical stellar library, and assuming the \cite{chabrier2003galactic} IMF. The synthetic template library contains $636$ SSPs unevenly spaced in linear age and [Z/H], covering $53$ age from $0.03$ Gyr to $14$ Gyr and $12$ metallicities from [Z/H]=$-2.27$ to [Z/H]=$0.4$.
We made the simple assumption that each region can be described by the combination of two SSPs, one older than $2$ Gyr and the other younger than $4$ Gyr. The old component represents the precollisional population of the original galaxy and is not expected to be younger than $2$ Gyr, while the young component can trace the possible stars recently formed as a consequence of the impact with the bullet companion. 
According to dynamical measures, the collision has taken place in the last few hundred Myrs, but we leave a larger age range (up to $4$ Gyr) as a prior, to be flexible towards a scenario where no stars are formed during the impact in some regions. In the latter case, the age range overlap between the two SSPs (old and young ones) allows us to take into account the possible complex SFH of the original galaxy that cannot be matched by a single SSP. 
We tested that the 2 SSPs model can successfully constrain the physical parameters of both populations until the young component (i.e. younger than $1$ Gyr) mixed with the older one (i.e. older than $3$ Gyr) represents a not-too-large fraction of the total mass (i.e. $< 30\%$). On the other hand, if the fraction of the young component is higher than $30\%$ of the total mass, the light contribution of this component will start to dominate the spectrum and the 2 SSPs model will start to fail to constrain the older population.

\begin{figure}
   \centering
   \includegraphics[width=0.5\textwidth]{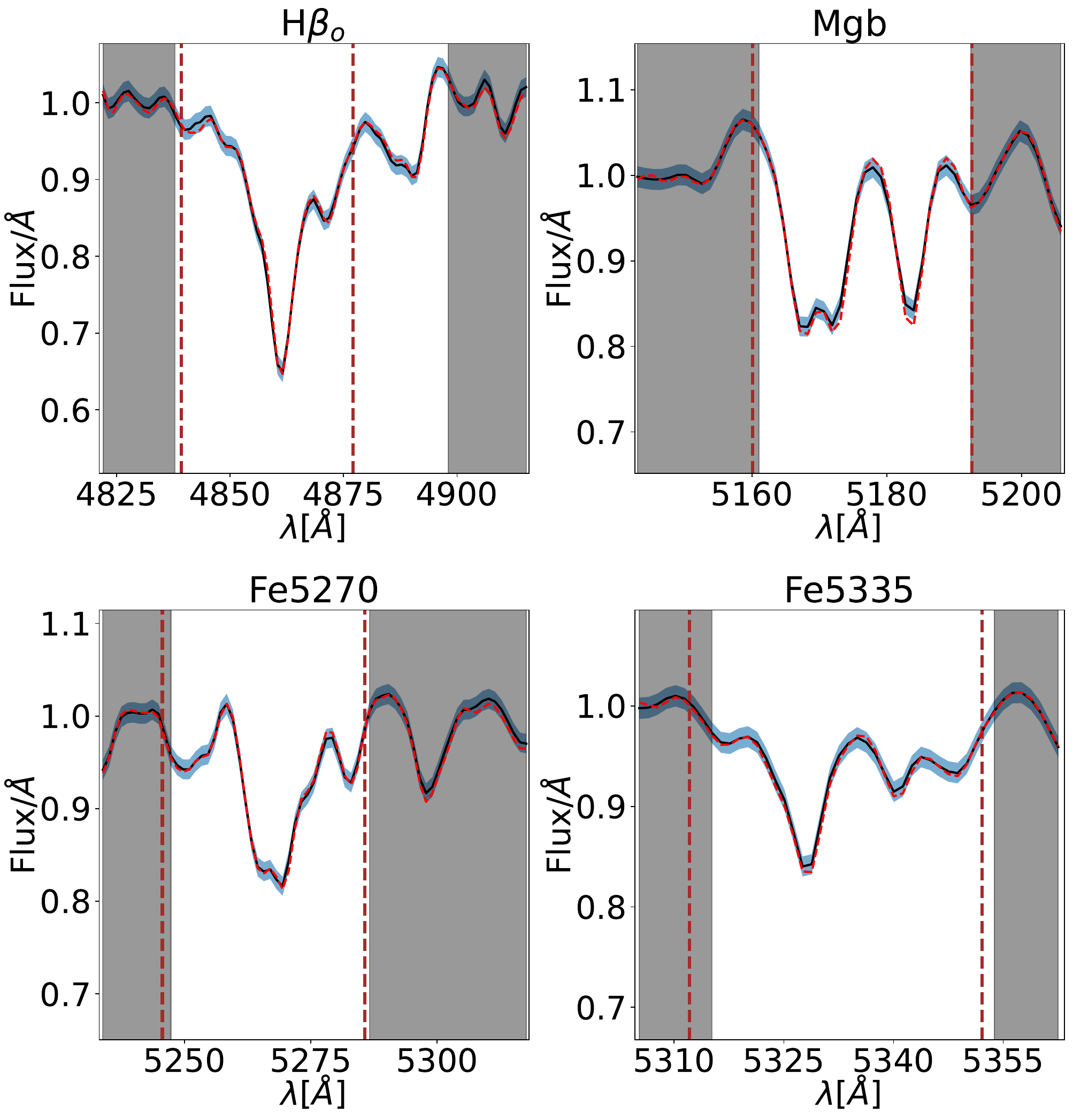}
      \caption{FIF application on the four spectral indices of the inner ring spectrum used. The vertical dashed brown lines mark the central feature of each index. The spectral features are normalised using the index pseudo-continua (grey shaded regions). The black lines and the blue shaded regions represent the observed spectrum and the relative uncertainties, respectively, while the superimposed red dashed lines are the best-fit obtained from the posterior distribution. 
              }
         \label{fig:fif}
   \end{figure}

We based the comparison between the stellar spectra and the synthetic spectral templates on the analysis of four spectral indices: H$\beta_{o}$ \citep{cervantes2009optimized}, sensitive to the age of the stellar populations, and Mgb, Fe$5270$ and Fe$5335$ \citep{worthey1994old}, sensitive to the stellar metallicity.
Given the high quality and resolution of the MUSE spectral data, we decided to follow the full-index fitting  approach \citep[FIF,][]{martin2014stellar}
to derive the stellar population parameters of each region of the Cartwheel galaxy. Differently from the more classic index fitting approach, with FIF the comparison of the flux within a specific absorption feature of an index (with respect to its continuum value) is made pixel by pixel and not on average. 
The comparison of single pixel fluxes within the index window is then more efficient in breaking the degeneracy between age and metallicity with respect to the classic index analysis, since not only the strength of the absorption feature is taken into account, but also its specific shape, which brings information on the stellar population parameters \citep[][]{martin2019fornax}. 
The FIF technique has several advantages also with respect to the full spectral fitting, because in the former case only the spectral features are fitted (not the continuum), optimising the information gained and reducing the computational time. Figure~\ref{fig:fif} shows an example of the application of the FIF approach on the inner ring of the galaxy. The best-fit template follows the shape of each spectral feature in an accurate way and then optimises the information gained pixel by pixel.

\subsection{Hierarchical Bayesian modelling}
\label{subsection:hierarc}
\begin{figure*}
   \centering
   \includegraphics[width=0.95\textwidth]{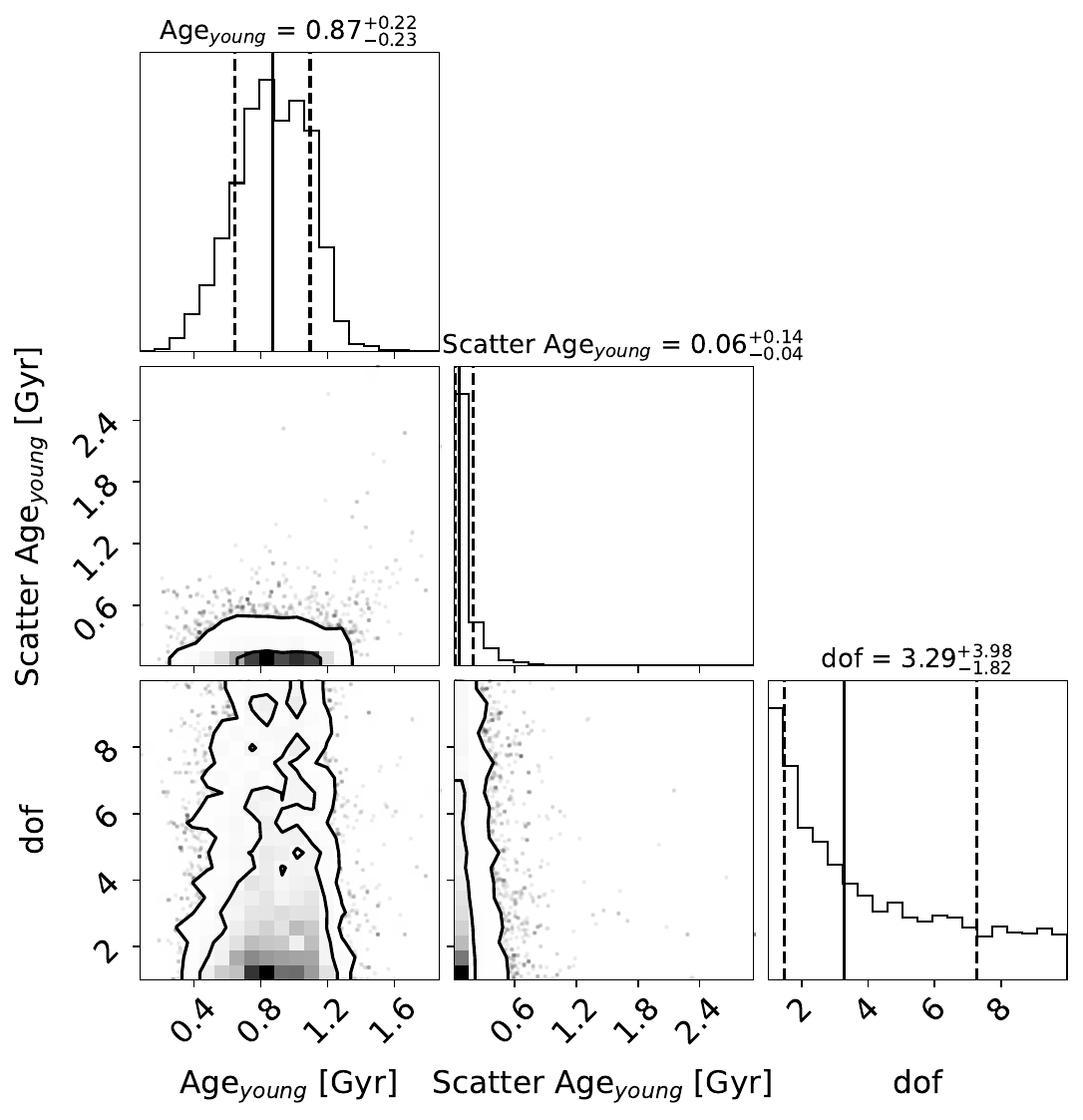}
      \caption{Corner plot summary of the posterior distributions of the Age$_{young}$, its scatter, and the degrees of freedom of the Student-t distribution obtained for the four subregions of the {inner ring} of the Cartwheel galaxy. The panels on the diagonal show the 1D histogram for each model parameter obtained by marginalising over the other parameters, with median and $16\%$ -- $84\%$  intervals indicated by solid and dashed lines, respectively. The off-diagonal panels show 2D projections of the posterior probability with contours at the $68\%$ and $95\%$ levels.
              }
         \label{fig:joint_innerring}
   \end{figure*}

In Section~\ref{section:specextract} we explained that, given the large dimensions of three out of the four regions selected in the Cartwheel galaxy, we could divide them in subregions in order to perform a spatially resolved analysis of the entire galaxy.
To assess the stellar population parameters of each region, we then need to combine the results obtained from their different subregions and we adopted a hierarchical bayesian modeling. In the hierarchical framework, our models consist of two levels, being the first one that of the individual measurements in each subregion, and the second one their distribution within the entire region. Following an `a posteriori' approach, as in \cite{beverage2023}, as the first level of the models we computed the posterior probability of age and metallicity of each SSP template used to fit the stellar spectra and their relative fraction for each subregion, using the likelihood given by $\mathcal{L} = e^{-\chi^2/2}$, with
\begin{equation}
    \chi^2 = \sum_{i}{\left(\frac{F_{obs_i}-F_{syn_i}}{\sigma_{obs_i}}\right)}^2
,\end{equation}
where F$_{syn_i}$ is the flux of the synthetic spectrum along the feature of each index, and $F_{obs_i}$ is the flux of the observed spectrum with the error $\sigma_{obs_i}$. The index $i$ indicates the $i$th pixel on which the calculation takes place.
We assumed a uniform prior for age and [Z/H] for both SSPs, and for the relative fraction of the two components, i.e. young and old. We considered an age range from $2$ Gyr to $14$ Gyr for the old component and from $0.03$ Gyr to $4$ Gyr for the young one, while we used the whole metallicity range for both the two components, from [Z/H]=$-2.27$ to [Z/H]=$0.4$. We explored the parameter space with a Markov Chain Monte Carlo method \citep[MCMC,][]{gilks2005m}, and the $68\%$ confidence interval is delimited by the $16^\circ$ and $84^\circ$ percentile.
Then, as a second level modeling, we fitted the posterior distribution of each parameter in each subregion using a Student-t distribution. This allows us to take into account the effects of a higher kurtosis in the distribution of the stellar parameters. 
With this approach, we can then obtain a mean value of the estimated parameters in the whole regions with a reliable estimate of their errors, accounting also for their intrinsic scatter among the different subregions. For this second fit, we assumed the same uniform prior described above for each stellar population parameter, then a logarithmically uniform prior for the intrinsic scatter in the range from $0.01$ to $10$, and a logarithmically uniform prior for the degrees of freedom of the Student-t distribution, in the range from $1$ to $10$, where $1$ is an heavy-tailed distribution and $10$ is a Gaussian one.
Figure~\ref{fig:joint_innerring} shows an example of the posterior probability distribution of the Age$_{\text{young}}$, its scatter and the degrees of freedom of the Student-t distribution for the inner ring.

Differently from the other regions, for the nucleus we analysed a single spectrum considering only the first of the two layers of the modeling. As it will be better explained in Sect~\ref{subsection:mcspf}, the analysis of the nucleus using the FIF approach did not reveal the presence of a density or shock wave induced past SF population (i.e. young component). Therefore for this region we adopted a simpler SFH described by a single SSP model. Similarly, the FIF analysis of all the subregions in the outer ring did not reveal the presence of the precollisional population (i.e. old component). Therefore,  also for this region we adopted the single SSP modeling, with the goal of finding not only a mean stellar age and metallicity but also their possible intrinsic scatter between the $16$ subregions in which we divided this region.

Table~\ref{table:valerr} reports the velocity dispersion, the median values of the marginalised posterior distribution of age and metallicity of the 2 SSPs and their relative mass fraction for each spatial region. Table~\ref{table:scatter} reports the median values of the marginalised posterior distribution of the scatter of age and metallicity of the 2 SSPs and their relative mass fraction for each spatial region. It is immediately clear that the scatter between the subregions within the inner ring and the in-between is negligible or comparable to the uncertainties for all the measured parameters. On the other hand, we measured a noticeable scatter on the mean age in the outer ring, which we discuss in detail in Sect.~\ref{subsection:outering}.

\begin{table*}
\caption{Median values of the marginalised posterior distribution for each subregion. }            
\label{table:valerr}      
\centering          
\begin{tabular}{l c c c c c c}     
\hline\hline       
Region & $\sigma_{\text{region}}$& Age$_{\text{old}}$ & [Z/H]$_{\text{old}}$ & Age$_{\text{young}}$ & [Z/H]$_{\text{young}}$ & Fraction$_{\text{old}}$\\
&[km s$^{-1}$]& [Gyr] & [dex] & [Gyr] & [dex] &  \\
\hline   \\                  
  nucleus & 72 & 3.17$^{+0.43}_{-0.34}$ & -0.06$^{+0.05}_{-0.05}$ & - & - & - \\\\      
   inner ring & 73 &5.62$^{+3.94}_{-1.97}$ & 0.20$^{+0.11}_{-0.13}$ & 0.87$^{+0.22}_{-0.23}$ & -0.87$^{+0.32}_{-0.26}$ & 0.94$^{+0.02}_{-0.03}$\\\\
   in-between & 65 & 3.42$^{+1.80}_{-0.90}$ & 0.12$^{+0.11}_{-0.11}$ & 0.80$^{+0.06}_{-0.14}$ & -1.06$^{+0.09}_{-0.07}$ & 0.84$^{+0.03}_{-0.03}$\\\\
   outer ring & 54 &- & - & 0.10$^{+0.02}_{-0.02}$ & -1.11$^{+0.04}_{-0.05}$ & -\\\\
\hline                  
\end{tabular}
\tablefoot{The first column shows the velocity dispersion measurement of each region obtained from the pPXF fit. The subsequent columns show the age and stellar metallicity of the two SSPs model and their relative fraction. The errors on the median values refer to the 16th$^{}$ and 84th$^{}$ percentiles. }
\end{table*}

\begin{table*}
\caption{Median values of the marginalised posterior distribution of the intrinsic scatter within the different subregions.}             
\label{table:scatter}      
\centering          
\begin{tabular}{l c c c c c}     
\hline\hline       
Region & Age$_{\text{old}}$ & [Z/H]$_{\text{old}}$ & Age$_{\text{young}}$ & [Z/H]$_{\text{young}}$ & Fraction$_{\text{old}}$\\
& [Gyr] & [dex] & [Gyr] & [dex] &  \\
\hline   \\                  
  nucleus  & N.A. & N.A. & - & - & - \\\\      
   inner ring  &0.14$^{+0.77}_{-0.11}$ & 0.03$^{+0.06}_{-0.02}$ & 0.06$^{+0.14}_{-0.04}$ & 0.05$^{+0.17}_{-0.04}$ & 0.02$^{+0.02}_{-0.01}$\\\\
   in-between  &0.06$^{+0.26}_{-0.04}$ & 0.03$^{+0.06}_{-0.02}$ & 0.03$^{+0.05}_{-0.01}$ & 0.04$^{+0.10}_{-0.03}$ & 0.02$^{+0.02}_{-0.01}$\\\\
   outer ring &- & - & 0.07$^{+0.02}_{-0.02}$ & 0.11$^{+0.06}_{-0.05}$ & -\\\\
\hline                  
\end{tabular}
\tablefoot{The columns show the scatter of age and stellar metallicity of the two SSPs model and their relative fraction. The errors on the median values refer to the 16th$^{}$ and 84th$^{}$ percentiles.}
\end{table*}

\subsection{Properties of the nebular emissions}
\label{subsection:nebular}

   \begin{figure*}
   \centering
   \includegraphics[width=0.95\textwidth]{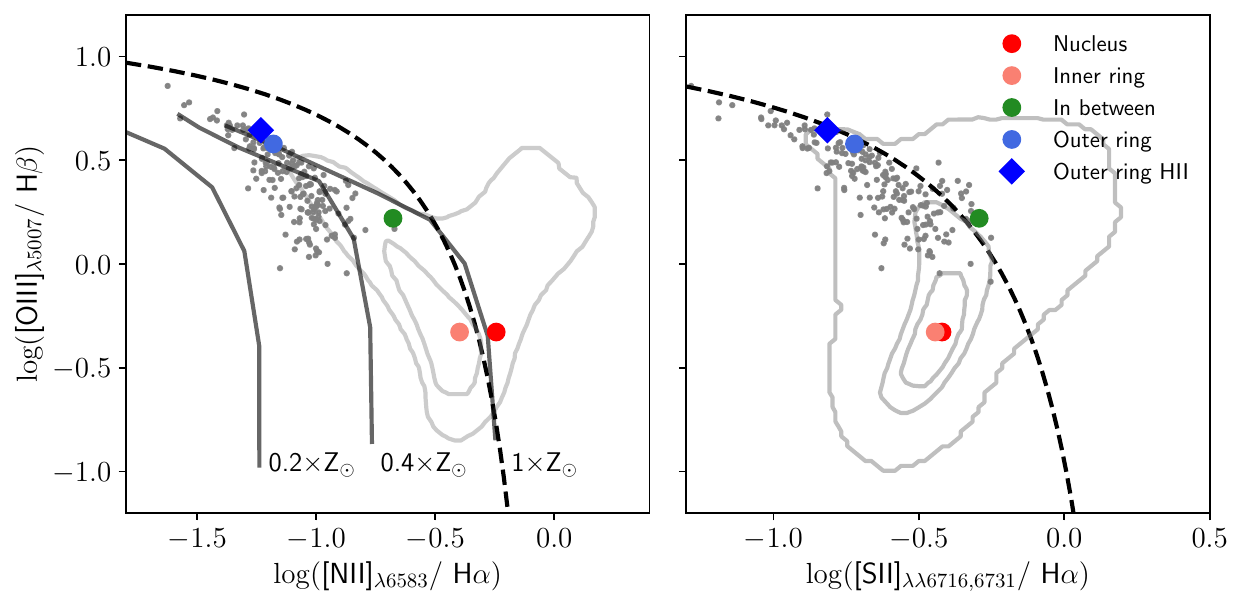}
      \caption{BPT diagnostic diagrams [OIII]/H$\beta$ vs. [NII]/H$\alpha$ (left), and [SII]/H$\alpha$ (right) derived from the composite spectra extracted from the five regions of the Cartwheel galaxy. The small grey points are the individual HII regions from \citet{zaragoza2022nebular}. The grey contours are obtained from the nuclear spectra of a sample of 50,000 SDSS galaxies at 0.03 < $z$ < 0.08 whose emission lines are detected at $S/N$> 5. The dashed lines indicate the separation between the stellar photoionisation processes (towards to left) and AGN/shock ionisation (top and right) as defined in \cite{kauffmann2003} (K03, left panel) and in \cite{kewley2006} (K06, right panel). The black solid lines in the left panel show the tracks from photoionisation models of \cite{kewley2001} for three different metallicities ($0.2$, $0.4$, and $1$Z$_\odot$).}

         \label{fig:bpt}
   \end{figure*}

The nebular spectrum contains valuable information to help in constraining the physical parameters of a galaxy, most notably the gas ionisation mechanism, its metallicity and the dust extinction in the stellar birth places. Figure \ref{fig:bpt} shows the BPT \citep[][]{baldwin1981} diagnostic line ratios [OIII]/H$\beta$ vs. either [NII]/H$\alpha$ (left panel), or [SII]/H$\alpha$ (right panel) derived from the composite spectra extracted from the four regions of the Cartwheel galaxy, and a fifth region constructed as the sum of the emission from the HII regions identified in the external ring in \citet{zaragoza2022nebular}. Both diagrams show that the ionisation of the {\it Outer ring} and the {\it in-between} regions is consistent with photoionisation by recent star-formation. In particular, the {\it Outer ring} exhibits a strong ionisation parameter, which is even stronger when we analyse the HII regions only, indicating significant star formation activity in the region. The template spectrum of the HII regions in the {\it Outer Ring} has line ratios in agreement with the locus of the points of the individual HII regions from \citet{zaragoza2022nebular}, as also confirmed by \cite{mayya2023}.
Conversely, the {\it Nucleus} has line ratios consistent with a weak active galactic nucleus (AGN) when the [NII] BPT diagrams is used, although the line ratios could also be consistent with hot evolved stars \citep{belfiore2016}. No AGN is detected with X-ray Chandra data, and only a variable source with maximum luminosity of L$_X$ ($2-10$ keV) $ = 1.4 \times 10^{39}$ erg s$^{-1}$ is present in the nuclear region \citep{Salvaggio2023}, suggesting the latter hypothesis. The {\it Inner Ring} has line ratios consistent with those of the {\it Nucleus}, indicating a similar origin of the nebular ionisation.

\begin{figure}
   \centering
   \includegraphics[width=0.5\textwidth]{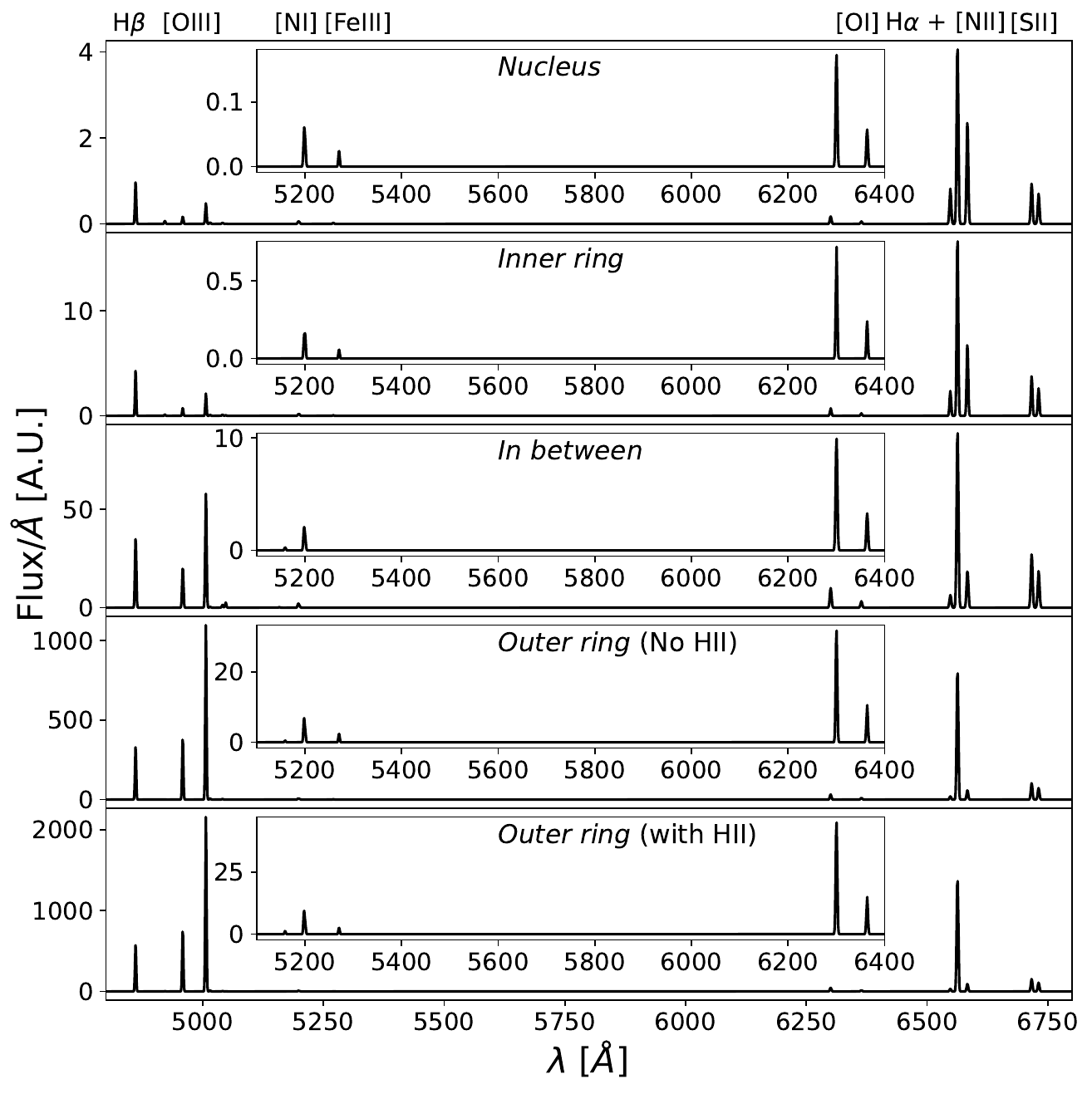}
      \caption{Nebular spectra of each region of the Cartwheel galaxy, as indicated in each panel. The inset shows a zoom onto the central part of the spectrum. Strong emission lines are reported at the top of the figure.
              }
         \label{fig:figneb}
   \end{figure}

Given the predominant role of stellar ionising photons on the ionisation conditions of the nebular gas, it is possible to derive the gas-phase metallicity from a combination of nebular emission line ratios, and to compare it with that of the two stellar populations adopted to model each region within the Cartwheel galaxy (see Section ~\ref{subsection:stellarcont}). 
Figure~\ref{fig:figneb} shows the nebular spectra extracted from each region. We noticed the similarity of the nebular spectra extracted from the nucleus and from the inner ring, implying a similar gas-phase metallicity. Moving towards the outer ring, the increase of the [OIII] lines over the $H\beta$ line and the corresponding decrease of the [NII]/$H\alpha$ ratio mark the evidence of a decrease of the gas-phase metallicity. 
From the empirical calibration of \cite{curti2017met} for strong-line diagnostics of gas-phase metallicity, we derived metallicity estimates based on R$_{3}$, N$_{2}$ and O$_{3}$N$_{2}$ indicators. The different estimates are in excellent agreement with each other. In Table~\ref{table:tneb} we reported the values resulting from the R$_{3}$ indicators, where the uncertainty has been derived from the dispersion of the corresponding empirical curve. As a further check, we also applied the empirical calibration by \cite{cl2001}, finding gas-phase metallicity values well matching the previous ones. From Table~\ref{table:tneb} we can see the decrease of the gas phase metallicity from the solar value in the inner part of the Cartwheel galaxy to less than half of the solar value in the outer ring.

We derived an estimate of the amount of dust extinction by measuring the Balmer decrement between H$\alpha$ and H$\beta$, assuming a case B value of their ratio equal to $2.86$ and the Calzetti attenuation law \citep{Calzetti:2001}. 
We then derived the current SFR in each region from the luminosity of the H$\alpha$ line, corrected for the calculated extinction, adopting the conversion factor of \cite{Kennicutt98} reported to the Chabrier IMF   \citep[][]{Driver2013}. Under the assumption of the H$\alpha$/SFR conversion factor of \cite{Kennicutt98} and the extinction curve of \cite{Calzetti:2001}, both SFR and A$_{V}$ are subject only to the statistical errors in measuring line fluxes, and since these are very small ($<0.5\%$), they result in negligible uncertainties in their estimates. 
Table~\ref{table:tneb} summarises all the quantities measured on the nebular component of each of the four regions.
\begin{table}
\caption{Nebular parameters.}             
\label{table:tneb}      
\centering          
\begin{tabular}{l c c c}     
\hline\hline       
Region & \ [Z/H]$_{gas}$ &  A$_{V}$ & SFR\\
\    & [dex]  & [mag] &  $\mathrm{M_\odot~yr^{-1}}$ \\
\hline   \\                  
  nucleus    & 0.01 $\pm \ 0.07$ & 2.46 & 0.13 \\\\    
  inner ring & 0.01 $\pm \ 0.07$ & 2.21 & 0.38 \\\\    
  in-between & -0.17 $\pm \ 0.07$ & 0.66 & 0.73 \\\\    
  outer ring (No HII) & -0.32 $\pm \ 0.07$ & 0.56 & 5.52 \\\\  
  outer ring (Only HII) & -0.41 $\pm \ 0.07$ & 0.47 & 3.73 \\\\  
  outer ring (With HII) & -0.36 $\pm \ 0.07$ & 0.52 & 9.17 \\\\   
\hline                  
\end{tabular}
\end{table}

\subsection{Reconstruction of the SFHs using MC-SPF}
\label{subsection:mcspf}
Finally, we also jointly fitted the photometric and spectroscopic data to extract the SFH of each region of the galaxy using the Python code Monte Carlo Spectro-Photometric Fitter \citep[MC-SPF,][]{fossati2018virgo}. As input data, we used the 1D spectra extracted in the four regions of interest, complemented by the fluxes in each photometric band integrated over the same regions, after subtracting an appropriate background for each region and band. The instrumental point spread function of diffraction limited observations and the seeing conditions of ground based data vary from $0.6$ arcsec for JWST-NIRCam F090W to $\approx5$ arcsec for the GALEX data. However, given the large apertures of spectral extraction adopted we did not match the images to a common PSF.

On the MC-SPF code side, we made numerous updates motivated by the needs of this investigation. First, the complex SFH of the different regions of the Cartwheel make the use of parametric SFHs impractical. We therefore chose a nonparametric SFH approach, informed by the stellar and nebular parameters obtained as described in Sect.~\ref{subsection:stellarcont} and~\ref{subsection:nebular}. The aim of this third approach is to verify how the ages recovered in the analysis based on the FIF for the stellar components can be produced by a simplified SFH. For each region, we defined a piece-wise constant SFH in three age bins, an old one ranging from the age of the old SSP to the age of the young SSP and characterised by the stellar metallicity of the old SSP; a young one spanning the time from the young SSP age to $30$ Myr ago (the maximum age of the ionising stars); and the youngest one, linked to the current star formation activity and to nebular line emission, ranging from $30$ Myr ago to the present day.  The last two bins have the stellar metallicity of the young SSP. The code performs the fit varying the relative weights of SFR in these three time bins, determining their relative fractions defined as
Frac$_{\text{old}}$, Frac$_{\text{young}}$ and Frac$_{\text{youngest}}$ (where Frac$_{\text{youngest}}$ = $1 -$ Frac$_{\text{old}}$ $-$ Frac$_{\text{young}}$), respectively. We generated model spectra from these SFHs using SSPs from the latest version of the \cite{bruzual2003stellar} \citep[named as C$\&$B, see][]{plat2019constraints,sanchez2022sdss}, where we selected the closest metallicity template to the one identified in each bin of the SFHs.
Each SSP template provides $220$ spectra computed at different time steps ranging from $0.01$ Myr to $14$ Gyr, with a metallicity ranging from $-1.7$ dex to $0.4$ dex.
We assumed a Chabrier IMF with $M_{UP} = 100 {\mathrm M_\odot}$  and the MILES stellar library, in order to be consistent with the models used for the retrieval of the stellar parameters in Sect.~\ref{subsection:stellarcont}.

For the nucleus and the outer ring, the FIF analysis shows that disentangling stellar populations of largely different ages is difficult as these spectra can be modelled with a single population of stars (old in the case of the nucleus and young in the case of the outer ring). 
For these reasons, we followed a different approach in building the SFHs for these regions. We generated a piece-wise constant SFH in only two age bins, with a single stellar metallicity and an upper age of $10$ Gyr. In addition to leaving the star formation fraction of the two bins variable, in the case of the nucleus the SFH is regulated by a free parameter (JumpAge Old) when the old component will cease to form. At a later time (JumpAge Young), the SF activity will resume. These models have three free parameters and are highly flexible in order to determine the old SFH and a possible rejuvenation through recent star formation. In the case of the outer ring, instead, JumpAge Old determines the time at which the first stars are formed, while JumpAge Young marks the time at which an increase (or decrease) of the star formation activity occurred and that level is kept until the time of the observation. We verified that, with the large flexibility allowed by these models, the assumption of an overall duration of the SFH of $10$ Gyr has no impact on the results presented in the following sections.


MC-SPF includes a modelling of emission lines using line ratios from \cite{byler2017nebular} scaled to the Lyman continuum luminosity of the stellar templates and assuming the gas-phase metallicity measured from emission line ratios. We also included dust attenuation with a \cite{Calzetti:2001} law, where the flux from stars older than $10$ Myr is attenuated with a curve normalised by a free parameter ($A_V$) in the fitting procedure, while the same curve is normalised by $2.27 \times A_V$ for stars younger than $10$ Myr. MC-SPF uses the MultiNest \citep{feroz2008multimodal,feroz2009multinest,feroz2019importance} code to sample the posterior distribution and obtain the best-fitting parameters with reliable uncertainties in a Bayesian framework. Results obtained with the MC-SPF code will be discussed in detail in the next Section.

   \begin{figure*}
   \centering
   \includegraphics[width=0.95\textwidth]{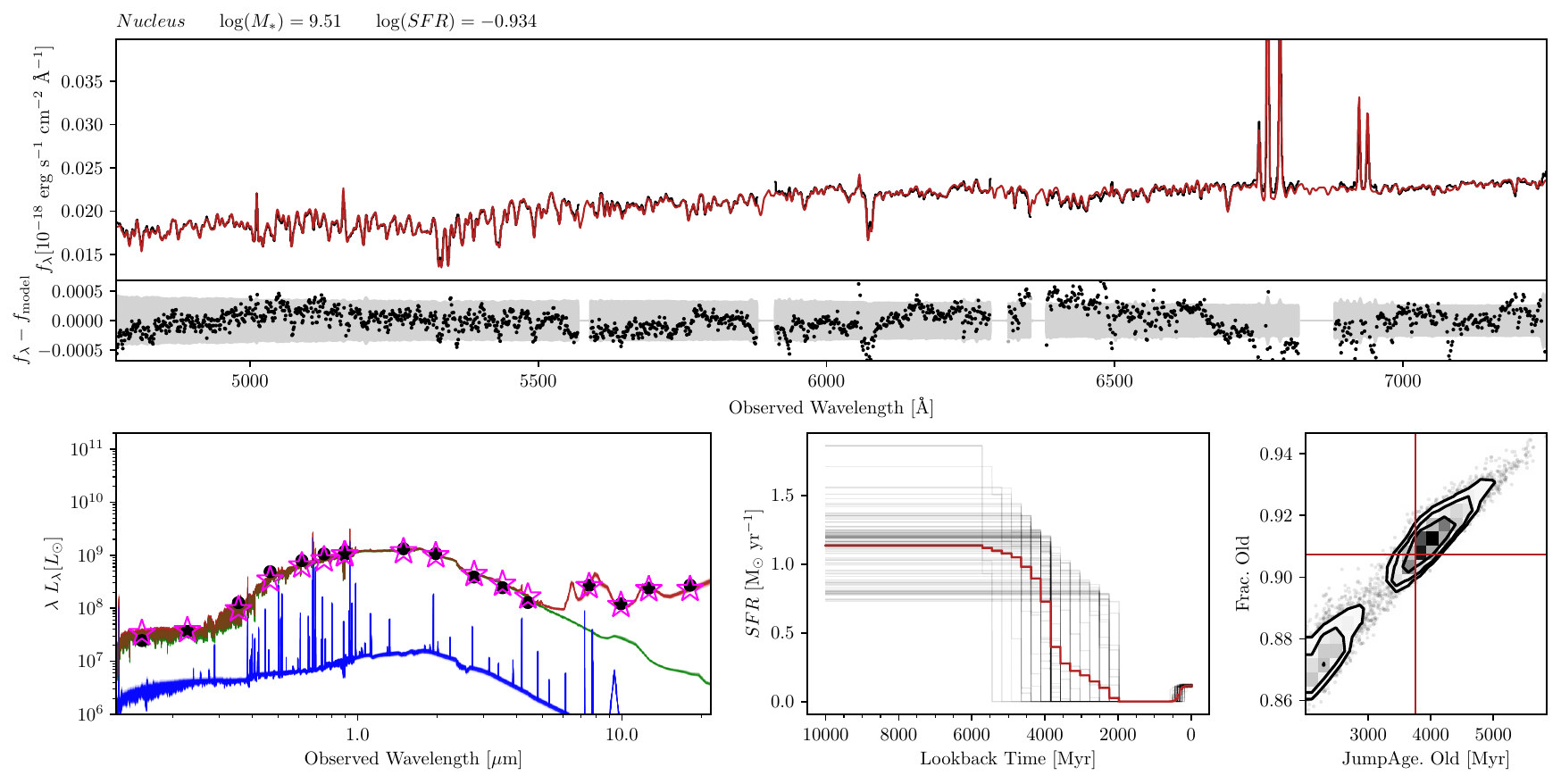}
      \caption{Results of the MC-SPF fitting for the nucleus region of the Cartwheel galaxy. Upper panel: MUSE spectrum (black) and best-fitting model (dark red). Spectral intervals dominated by bright skylines are not used in the fit. Fit residuals (Data-Model) are shown below the spectrum and the grey shaded area shows the $1\sigma$ uncertainties. Lower left panel: Photometric data points in black. The dark red lines are the total model including the dust emission, dominating at $\lambda > 5\mu$m, while the blue (green) lines show the contribution of the stellar continuum and nebular line emission from young (old) stellar populations (Age $<>30$ Myr). The open magenta stars show the photometric datapoints from the best fit model. Different lines are $100$ random samples of the posterior distribution. Lower middle panel: Reconstructed SFH from the fitting procedure. The dark red solid line is the median of the SFH samples. Lower right panel: Marginalised likelihood maps for the JumpAge Old and Frac. Old fit parameters. The red lines show the median value for each parameter, while the black contours show the $1$, $2,$ and $3\sigma$ confidence intervals. 
              }
         \label{fig:mcspf}
   \end{figure*}

   \begin{figure*}
   \centering
   \includegraphics[width=0.95\textwidth]{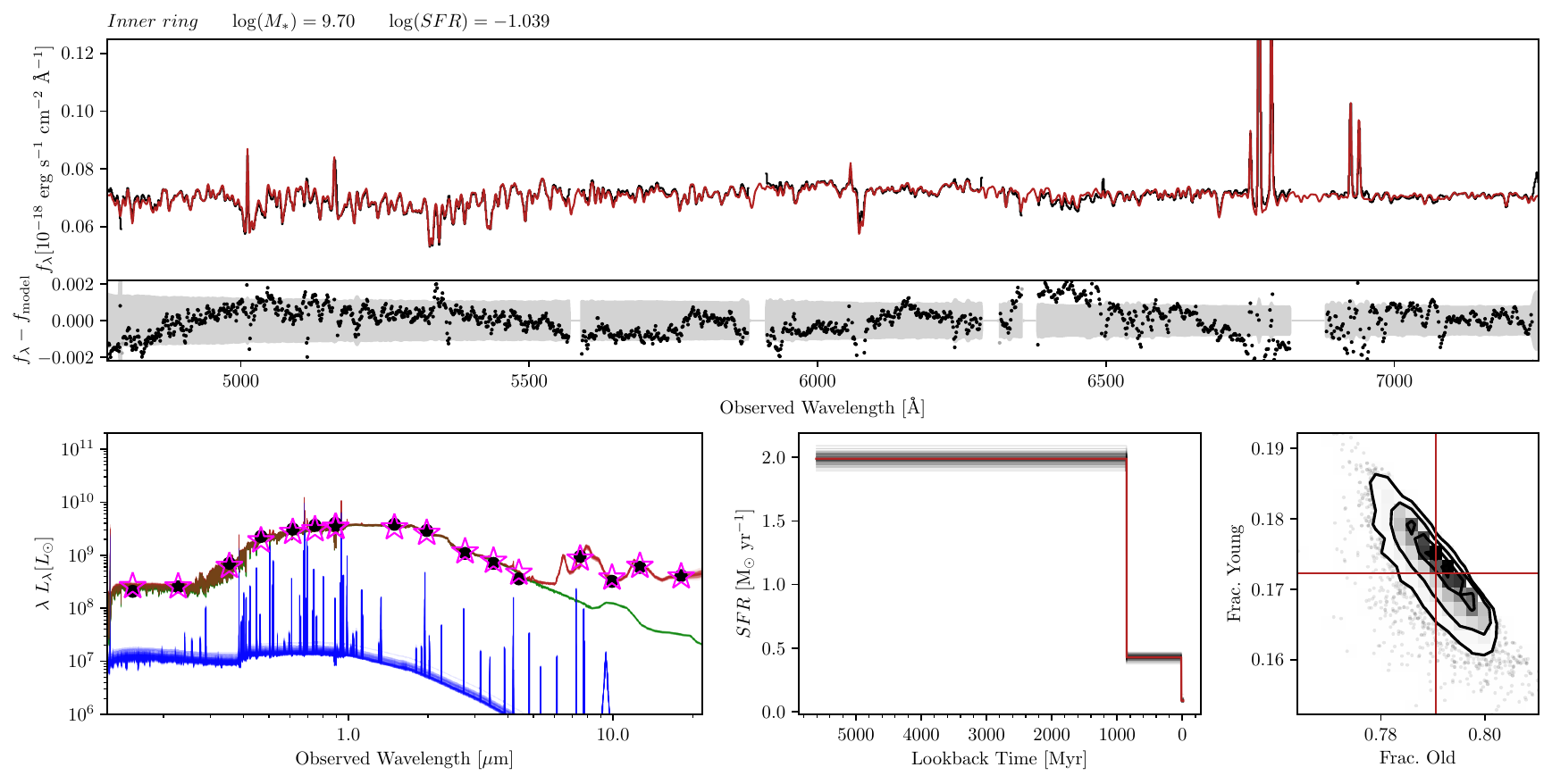}
      \caption{Same as Fig.~\ref{fig:mcspf} but for the {inner ring} of the Cartwheel galaxy. In this case, the lower right panel shows the marginalised likelihood maps for the Frac. Old and Frac. Young fit parameters.
              }
         \label{fig:mcspf_inner}
   \end{figure*}

      \begin{figure*}
   \centering
   \includegraphics[width=0.95\textwidth]{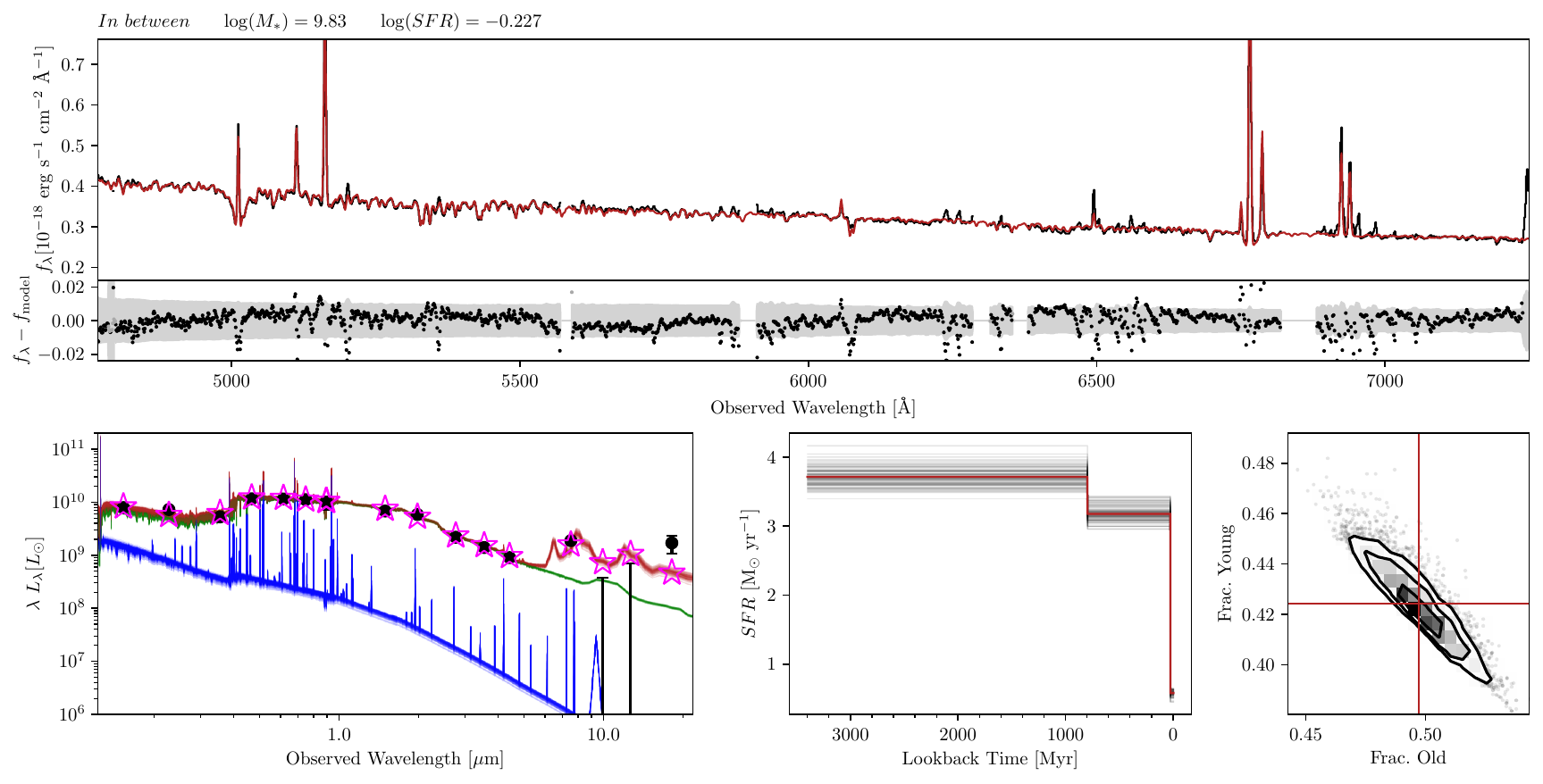}
      \caption{Same as Fig.~\ref{fig:mcspf} but for the {in-between} region of the Cartwheel galaxy. In this case, the lower right panel shows the marginalised likelihood maps for the Frac. Old and Frac. Young fit parameters.
              }
         \label{fig:mcspf_in_between}
   \end{figure*}
   
\section{Results}
\label{section:results}
As detailed in the previous sections, we extracted the spectra corresponding to the four spatial regions shown in Fig.~\ref{fig:cartwheelha} and we analysed them in their stellar and nebular components. We also derived the SFH for each region of the Cartwheel galaxy (see Section ~\ref{subsection:mcspf}). In the following we present the results obtained. In particular, we discuss the ages of the stellar components derived for each separate region via the FIF technique in the framework of the SFH derived by means of the MC-SPF analysis, and we compare the stellar metallicities with the nebular ones. 

\subsection{Nucleus}
As anticipated in Sect.~\ref{subsection:hierarc} and~\ref{subsection:mcspf}, our analysis using the FIF approach did not reveal the presence of a density or shock wave induced (post-collisional) population. Therefore we adopted a single SSP model to obtain a mean value of stellar age and metallicity of the nucleus. We found that this region is dominated by a stellar component of $\sim 3.2$ Gyr with a slightly subsolar metallicity (see Table~\ref{table:valerr}).
The gas-phase metallicity estimate (i.e. $0.01$ dex) is in broad agreement with that of the stellar component. The measured active SFR is compatible with zero, also given the fact that  we cannot exclude a contribution from a weak AGN to the emission lines. 
Figure~\ref{fig:mcspf} shows the results obtained from the MC-SPF code for the nucleus.
The derived SFH describe this region as mainly made during a star formation episode (modelled with a constant rate) that terminated $\sim 3.7$ Gyr ago, yielding about $91\%$ of the region stellar content, as can be seen in the middle bottom panel of Figure~\ref{fig:mcspf}, showing the posterior samples of the reconstructed SFH. In the same panel, the red line shows the average SFH which indicates an abrupt reduction in the star formation activity between 3 and 5 Gyr ago. The SFR has reduced below $50\%$ of its original value $\sim 4$ Gyr ago, in excellent agreement with the analysis of the JumpAge Old parameter. 
Even in the MC-SPF analysis, the input models do not include AGN contributions, and therefore the emission lines can only be fit with a very young population of stars accounting for $\approx 9\%$ of the total mass. Given the possible presence of a weak AGN, this value should be taken as an upper limit, which turns into a lower limit for the older populations of stars. 


\subsection{Inner ring}
Differently from the nucleus, in the inner ring we found a young component (i.e. $\sim0.8$ Gyr) albeit in a very small fraction (i.e. $6\%$) of the total mass, over a precollisional population with age of $\sim 5$ Gyr. This is reflected on the slightly higher active star formation (i.e. $0.38 \ \text{M}_\odot$ yr$^{-1}$) than the nucleus.
Figure~\ref{fig:mcspf_inner} shows the results obtained from the MC-SPF code for the this region.
The SFH representation matches the results obtained from the stellar analysis, producing about $79\%$ of the stellar content in a first $5$ Gyr of activity, which terminated roughly $1$ Gyr ago. We found a slightly super-solar stellar metallicity for the old component, which is consistent with the gas-phase one. 
Conversely, the metallicity of the young stellar SSP is estimated to be $~\approx 15\%$ of the solar value. Taken at face value, this could imply a more recent star formation event from metal poor gas, possibly related to the galaxy collision. However, the small mass fraction of the young SSP combined with the posterior distributions shown in Figure \ref{fig:joint_innerring} cannot rule out with high statistical significance that the younger stellar component is more metal enriched, reaching a significant fraction of the solar value. 

  
\subsection{In-between region}
In the {in-between} region we found a more consistent young component ($0.8$ Gyr, similar to what found in the inner ring), characterised by a sub solar stellar metallicity, representing $16\%$ of the total mass of the region. The gas-phase metallicity is measured as $67\%$ of the solar value.
Figure~\ref{fig:mcspf_in_between}, showing the results obtained from the MC-SPF code for this region, confirms the higher active SFR measured by the nebular analysis and the detection of a recently induced star formation activity beside the precollisional component. The different relative fractions of the two components obtained in the FIF analysis (i.e, more than $70\%$ of mass contained in the old component) with respect to those obtained in the MC-SPF (i.e. about $40\%$ of mass contained in the young component) could be due to the weakness of the FIF approach in reproducing the correct stellar mix on the basis of the simplified assumed SFH (two SSPs) and the spectral properties observed in a short wavelength range. Indeed, in the in-between the young component contains a relevant fraction of the total stellar mass (i.e. higher than $\sim 20\%$) and, given its young age, it dominates the stellar light emission, making it very hard to disentangle the properties of the two components on the basis of the estimates of few optical indices. In particular, on the basis of the stellar spectrum analysis, we obtain reliable measurements of the age and metallicity of the young component (i.e. the one dominating the optical emission), while those relative to the old one and their relative mass fraction are quite sensitive to the adopted SFH. The MC-SPF approach, due to its stronger assumption on the shape of the SFH and the larger wavelength coverage, is probably more reliable and successful in detecting both components and in deriving their characteristics. However, even if the overall picture is clear, only a physically motivated SFH from theoretical predictions on the origin of the old stars can ultimately shed light on the exact age and metallicity of the stellar populations in this region. 

\subsection{Outer ring}
\label{subsection:outering}
The outer ring emission is dominated by the most recent events, revealing a largely different picture with respect to that of the other regions.  As already noted in Sect.~\ref{subsection:hierarc}, the analysis of the optical spectrum, which is dominated by the youngest luminous stars, cannot reveal the possible presence of the precollisional component. For this reason, we further simplified the SFH used to represent this region assuming a single SSP. From the analysis of the optical stellar spectrum, we found that the outer ring has a contribution from stars older than $30$ Myr in addition to the current population powering the nebular ionisation. In particular, we found a noticeable scatter in the age of the nonionising population, ranging between $30$ and $170$ Myr in different subregions of the outer ring, with a mean age of $100$ Myr and $0.1Z_\odot$ stellar metallicity. In all regions of the outer ring, the contribution of the ionising stars (Age $< 30$ Myr) to the optical continuum is below $15\%$, providing indirect evidence of the inhomogeneity of this region. However, given the simplified SFH selected for this region in the FIF approach, we should be careful in interpreting this result. Indeed, a single SSP gives information about the light-weighted mean stellar population parameters that populate the considered region. In the particular case of the outer ring, the flux of the very young stars dominate the spectral region considered in the FIF approach, meaning that we should take this average range as a lower limit for the age of the oldest stars. Regarding the metallicity estimate, the measured value of about $10\%$ of the solar value has to be considered, like the age, a lower limit given that the gas-phase metallicity is about $0.4Z_\odot$.

\begin{figure*}
   \centering
   \includegraphics[width=0.95\textwidth]{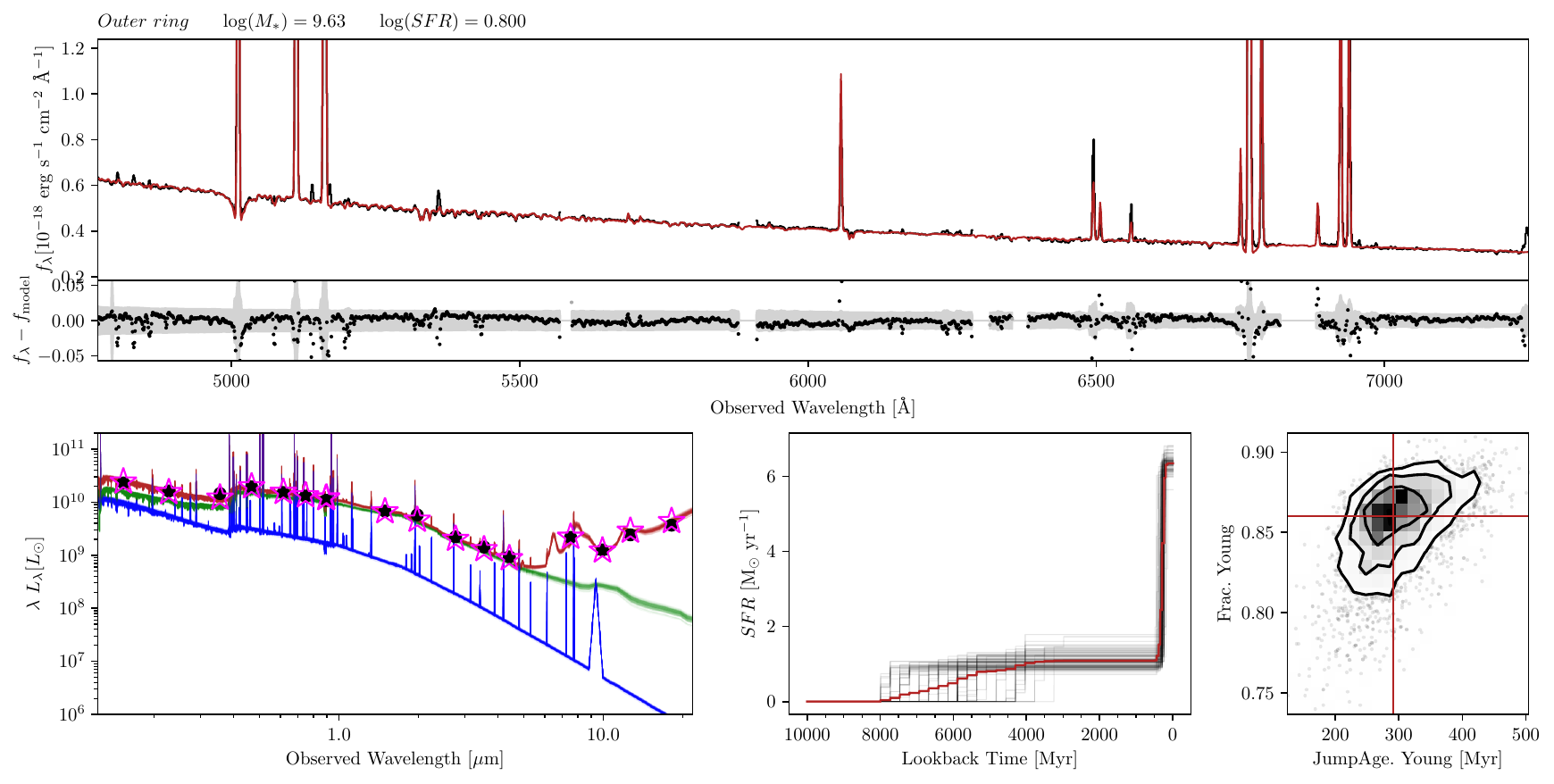}
      \caption{Same as Fig.~\ref{fig:mcspf} but for the {outer ring} of the Cartwheel galaxy. In this case the lower right panel shows the marginalised likelihood maps for the JumpAge Young and Frac. Young fit parameters.
              }
         \label{fig:mcspf_ext_ring}
   \end{figure*}

A complementary tool for understanding the stellar content of the outer ring is the use of the MC-SPF approach, which takes advantage of a larger wavelength baseline thanks to the joint fit of the photometry and the spectrum and to the more flexible SFHs. As explained in Sect.~\ref{subsection:mcspf}, differently from the other regions, in the outer ring we used a piece-wise constant SFH in two age bins, fitting a SFH model with three free parameters: the relative fraction of the two components, the lookback time for the onset of the SF creating the oldest stars, and the lookback time at which the star formation level has jumped from the old to the young component. We stress that, despite its simplicity, we chose this SFH to allow for a great degree of flexibility, in order to capture possible sudden variations in the SF activity that can be expected from the current starbusty activity of the outer ring. Figure~\ref{fig:mcspf_ext_ring} shows the results obtained with this code in the outer ring. We found that this region is largely dominated by young stars ($> 85\%$ of the mass) formed in a recent star formation event that, in our parametrisation, started around $300$ Myr ago. We also found a small ($< 15\%$) fraction of a  precollisional stellar component in this region of the galaxy. This component includes stars as old as $8$ Gyr, although the posterior samples shown in the SFH panel in Figure~\ref{fig:mcspf_ext_ring} indicate that the data can be reproduced also with stars not older than $4$ Gyr. As a result, there is a decreasing probability that stars older than $4$ Gyr are required by the model, possibly also due to the difficulty in detecting a small fraction of such old stars embedded in a dominant young population.

\section{Discussion}
\label{section:finaldiscussion}
The study of the stellar content in four regions of the Cartwheel galaxy at increasing radial distance from the centre allows us to define a possible framework to describe the recent evolution of this galaxy. 
We inferred a precollisional component with a mean age between $3$ and $8$ Gyr over all the galaxy with the only exception of the outer ring, which is dominated by stars formed in a starburst  that started roughly $300$ Myr ago and is still ongoing.
Even in the inner ring, which supposedly formed as a consequence of the impact suffered by the Cartwheel galaxy, we found that more than $90\%$ of the total mass is part of the precollisional component  (i.e. with an age of about $5-6$ Gyr), with the same gas-phase metallicity of the nucleus, suggesting that it belongs to the original galaxy. The slight increase in the star formation activity (see Table~\ref{table:tneb}) we detected in the inner ring with respect to the nucleus (confirmed also by the MC-SPF analysis, see Fig~\ref{fig:mcspf} and~\ref{fig:mcspf_inner}) well matches the expected SFR profile of a spiral galaxy of log(M/M$_\odot$) $= 10.7$ \citep{zaragoza2022nebular}
and it corresponds to a slight decrease of the gas-phase metallicity, as expected by the fundamental metallicity relation \citep[FMR,][]{mannucci2010}. 
The same trend is confirmed in the {in-between} region, where the SFR is more significant and the gas-phase metallicity $0.18$ dex lower than in the nucleus. Considering an average distance of the in-between region from the nucleus of about $10$ kpc, we derive a gas-phase metallicity gradient of $0.018$ dex/kpc, in perfect agreement with the gradient found by \cite{zaragoza2022nebular} derived from the analysis of individual HII regions identified from the nucleus towards the external ring. In \cite{zaragoza2022nebular}, the areas 4-5-6 in their Figure 11 and 12 correspond to our in-between region, and their estimate of the corresponding gas-phase metallicity is around $65\%$ of the solar value, well matching our estimate (see Table~\ref{table:tneb}). We found a good agreement also for the metallicity estimated in both the nucleus and the inner ring, where we measured a value around the solar one to be compared with an estimate of $80\%$ of the solar value derived from \cite{zaragoza2022nebular}.  

Considering the age of the stellar content, the main stellar component populating the innermost region of the Cartwheel galaxy up to the limit of the outer ring appears to be almost coeval (within 1$\sigma$ error), with little to no sign of very recent activity. A possible very young component formed in the last hundreds of Myrs due to the recent impact is almost totally absent in the nucleus and inner ring while it represents only about 20\% of the total mass in the in-between region.  
All these elements taken together suggest that, apart from the peculiar morphology, a large fraction of the stars in the Cartwheel galaxy are not affected by the recent impact with the companion bullet and the galaxy keeps the characteristics of a typical spiral galaxy in terms of age, metallicity and SFR distributions.

Differently from the other regions, in the outer ring we found mostly young stars (i.e. with ages $< 400$ Myr).
The estimated gas-phase metallicity in the outer ring considering all the flux coming from this region is $0.19$ dex lower than the value obtained in the in-between region, which is about $12$ kpc apart; this is consistent with the previously calculated gradient. 
This result is at odds with what \cite{zaragoza2022nebular} report for the gas-phase metallicities of the individual HII regions identified in the ring, which are lower than expected from the extrapolation of the inner gradient. Indeed, the mean gas-phase metallicity we measure in the outer ring is $\approx 45\%$ of the solar value, while the estimate obtained by \cite{zaragoza2022nebular} in the external ring HII regions is around $20\%$ of the solar value. The difference between the results obtained in the two works can be due to the different analysis we performed,
that averages over all the emitting outer ring area rather than being weighted towards the HII regions.
To test this hypothesis, we derived the gas-phase metallicity from the spectrum extracted by summing fluxes corresponding to the HII regions in the outer ring spectrum as defined by \cite{zaragoza2022nebular}. Indeed, we found that the HII regions selected by \cite{zaragoza2022nebular} in the external ring of the Cartwheel galaxy are less metal rich than the average gas contained in the outer ring, being the former less than $40\%$ of the solar value, i.e. $0.1 \ \mathrm{dex}$ lower than in the outer ring excluding the strong HII regions (see Table~\ref{table:tneb}). This result supports the idea outlined by \cite{zaragoza2022nebular} that the infall of metal poor gas from the intergalactic medium triggers strong star formation activity while diluting the metallicity of the existing gas that is still detectable outside the stronger HII regions.
It is worthy to note that the metallicity calibration curves adopted by \cite{zaragoza2022nebular} (i.e. \citealt{Pilyugin2016}) tend to give lower estimates than those we adopted (i.e. \citealt{curti2017met}) in particular in the lower metallicity regimes, increasing the discrepancy between our results and those by \cite{zaragoza2022nebular}.

On the other hand, the presence of such a high fraction of very young stars in the outer ring suggests that this is the only region dominated by very recent and ongoing star formation events triggered by the impact with the bullet companion galaxy. 
\cite{MayyaAge}, on the basis of Astrosat/UVIT and FUV imaging data, found evidence of a wide range of ages in the ring but still younger than $150$ Myr, suggesting a SFH that can be described by an average steady SFR of $ \approx 5 \ \mathrm{M_\odot~yr^{-1}}$ over the past $150$ Myr, with a possible, but less statistically significant, increase to $\approx 18 \ \mathrm{M_\odot~yr^{-1}}$ in the last $10$ Myr. 
Our results for the outer ring support the picture that it formed stars at an elevated rate in the recent past, with the onset of the recent starburst around $300$ Myr ago, and with an average rate of about $6.3$ $\mathrm{M_\odot~yr^{-1}}$. 
Our estimate of the SFR in the full {\it Outer ring} region, obtained from nebular emission lines, gives $9.25$ $\mathrm{M_\odot~yr^{-1}}$, in good agreement with the range of estimates by \cite{MayyaAge}.
Regarding the recent history of this region, our result implies the possibility that the actual external ring contains stars formed during the expansion of the collision wave, and that the starting time of the wave propagation happened about $300$ Myr ago, while the study of \cite{MayyaAge}, taking into account the brighter HII regions in the external ring only, finds a more recent value for the starting time of the wave propagation, i.e. around $150$ Myr. Nevertheless our estimates of the parameter $D = (L_{\rm young}+L_{\rm old})/L_{\rm young})= 6.67$, as first introduced in \cite[][see their Figure $7$]{MayyaAge}, and the EW(H$\alpha$)$= 190 \ \AA$ are in close agreement with the results obtained in the individual HII regions by \cite{MayyaAge}, providing further support to our modelling framework.
Our result well matches that of \cite{higdon1996}, who investigated the distribution and kinematics of the neutral hydrogen content in the Cartwheel galaxy. On the basis of the HI velocity field analysis, and assuming a constant expansion rate, they found that the impact took place around $300$ Myr ago. 
Our result is also consistent with the \cite{amram1998} work. Using the H$\alpha$ emission map obtained with a scanning Fabry–Perot interferometer, they dated the impact in the range from $210$ to $720$ Myr ago.

Apart from details on the timescale of the star formation events, we support the same picture presented in \cite{MayyaAge} and \cite{zaragoza2022nebular}, suggesting that the collision wave, while moving towards the actual position, drags the already formed stars sweeping the inner areas, as predicted by the recent collision model  by \cite{renaud2018}. 
Conversely, we describe the external ring as produced by a collision wave started not earlier than $400$ Myr ago, while \cite{renaud2018} predict a shorter lived ring lasting less than $200$ Myr. Furthermore, according to \cite{renaud2018} the ring should start to lose material during its expansion, material that is predicted to gravitationally fall back towards the nucleus within what we call the in-between region, triggering fresh star forming events. Our results derived from the analysis of the nucleus, inner ring  and in-between regions do not appear to support this picture (at least over the predicted short timescales) because we found that all the stellar populations in the Cartwheel galaxy follow the distribution expected in a disc galaxy, with the unique exception of the external ring.


\section{Summary and conclusions}
We present a novel analysis of the stellar population properties of the entire Cartwheel galaxy. Our study is based on the publicly available high-spatial-resolution spectroscopic data obtained with the MUSE instrument during the Science Verification observations. Photometric imaging covering the wavelength interval from the FUV to the IR region, including the recently available JWST/NIRCam and JWST/MIRI data, complement the analysed data set.
We divided the galaxy into four separate spatial regions, namely the `nucleus', the `inner ring', the `in-between region' and the `outer ring'. Given the large spatial dimensions of the selected regions (with the exception of the nucleus), we further divided each one into equally extended subregions in order to perform a spatially resolved analysis of the entire galaxy. For each of the four regions, we then derived the main physical parameters of their stellar content. We followed a two-step approach: First, (i) we compared four spectral indices sensitive to age and metallicity of the stellar populations (i.e. H$\beta$, Mgb, Fe$5270,$ and Fe$5335$) with model predictions, and  then (ii) we performed a full spectral fitting of the whole MUSE spectra combined with all the available photometric information to derive the SFH of each region. 

In the first step, we adopted the FIF procedure, which takes into account not only the strength of the absorption features but also their shape. Furthermore, we assumed that each region can be described by the combination of the precollisional population of the original galaxy with a recently formed population as a consequence of the impact with the bullet companion. 

In the second step, to extract the SFH of each defined region of the galaxy, we also jointly fitted the photometric and spectroscopic data using the MC-SPF code, updated with new features and SFHs tailored to this specific study. 
Our findings can be summarised as follows:

 \begin{itemize}
     \item  We find a dominant old component with a mean age of between $3$ and $8$ Gyr over all the galaxy with the only exception being the {outer ring}, which is entirely dominated by stars formed no earlier than $400$ Myr ago. Light from these nonionising stars dominates the observed optical continuum flux in the {outer ring} in spite of the extremely bright nebular lines.
     \item Apart from the peculiar morphology, a large fraction of the Cartwheel galaxy (from the {nucleus} to the in-between region) is not affected by the recent impact with the companion bullet and retains the characteristics of a typical spiral galaxy in terms of age, metallicity, and SFR distributions.
     \item Our analysis suggests a picture in which the collision shock wave, while moving towards the {outer ring}, sweeping the inner areas and dragging the already formed stars along with it,  as predicted by the recent collision model by \cite{renaud2018}, albeit on longer timescales.
     \item The lack of very young stars in the {inner ring} and in-between region does not match the \cite{struck1996}  and \cite{renaud2018} predictions of star forming material falling back towards the nucleus along the spokes.
 \end{itemize}

Our work demonstrates that the study of stellar population distributions, coupled with a nebular analysis, can provide information about the recent collision suffered by the Cartwheel galaxy. Our results show how optical spectroscopy and multi-wavelength photometry can precisely reconstruct the history of complex objects like RiGs. Existing and future facilities, such as IFU spectrographs on $30$ m telescopes, can extend this analysis beyond the local Universe, paving the way for statistical studies of RiGs.  

\begin{acknowledgements}
We thank the reviewer Y.D. Mayya for his comments that greatly improved the quality of the paper. We thank J.Trevor Mendel for the co-developement of the MC-SPF code, Emanuela Pompei for useful discussions, and Guido Consolandi, Andrea Bianco and Marco Landoni for the early development of this project. This work is partially based on public data released from the MUSE commissioning observations at the VLT Yepun (UT4) telescope under Programmes ID 60.A-9100(A), 60.A-9100(B) and 60.A-9100(C). This work is based in part on observations made with the NASA/ESA/CSA James Webb Space Telescope. The data were obtained from the Mikulski Archive for Space Telescopes at the Space Telescope Science Institute, which is operated by the Association of Universities for Research in Astronomy, Inc., under NASA contract NAS 5-03127 for JWST. These observations are associated with program ERO PID 2727. The authors acknowledge the ERO proposing team for developing their observing program with a zero-exclusive-access period.    
\end{acknowledgements}

%
%

\bibliographystyle{aa}
\bibliography{biblio}

\end{document}